\documentstyle[11pt]{article}
\newcommand{\nl}{\vskip 0.3cm \noindent}

\newcommand{\M}{{\bf M}({\bf K};{\bf X})}
\newcommand{\Ma}{{\bf M}({\bf K}_\alpha;{\bf X})}
\newcommand{\Mb}{{\bf M}({\bf K}_\beta;{\bf X})}
\newcommand{\MT}{{\bf M}_T({\bf K}_T;{\bf X})}

\newcommand{\Ml}{{\bf M}_L}
\newcommand{\ML}{{\bf M}_L({\bf K}_T;{\bf X})}
\newcommand{\be}{\begin{equation}}
\newcommand{\ee}{\end{equation}}
\newcommand{\ba}{\begin{eqnarray}}
\newcommand{\ea}{\end{eqnarray}}
\newcommand{\ga}{{\bf g}_\alpha}
\newcommand{\gb}{{\bf g}_\beta}
\newcommand{\la}{\lambda_\alpha({\bf K}_T;{\bf X})}

\newcommand{\gba}{{\bar{\bf g}}_\alpha}
\newcommand{\gbb}{{\bar{\bf g}}_\beta}
\newcommand{\fa}{{\widehat{\bf f}}_\alpha}
\newcommand{\fb}{{\widehat{\bf f}}_\beta}
\newcommand{\fba}{\widehat{\overline{\bf f}}_\alpha}
\newcommand{\fg}{{\widehat{\bf f}}_\gamma}
\newcommand{\fbr}{\widehat{\overline{\bf f}}_\rho}
\newcommand{\xmx}{{\bf X}-{\bf X}^{\prime}}
\newcommand{\xx}{{\bf X}^{\prime}-{\bf X}}
\newcommand{\xp}{{\bf X}^{\prime}}

\newcommand{\x}{{\bf X}}
\newcommand{\dt}{\frac{\partial}{\partial t}}
\newcommand{\dx}{\frac{\partial}{\partial\x}}
\newcommand{\dkt}{\frac{\stackrel{\leftarrow}{d}}{d\kt}}
\newcommand{\kk}{{\bf K}^{\prime}-{\bf K}}
\newcommand{\ja}{\stackrel{\leftrightarrow}{\partial}}
\newcommand{\jaa}{\stackrel{\leftrightarrow}{\partial_\alpha}}

\renewcommand{\sp}{{\stackrel{\leftarrow}{\partial}}}
\newcommand{\spa}{{\stackrel{\leftarrow}{\partial_\alpha}}}

\newcommand{\epha}{e^{\phi_\alpha(\x)}}
\newcommand{\ephg}{e^{\phi_\gamma(\x)}}
\newcommand{\ep}{{\bf\varepsilon}({\bf K};\x)}
\newcommand{\epa}{{\bf \varepsilon}({\bf K}_\alpha;\x)}
\newcommand{\kt}{{{\bf K}_T}}
\newcommand{\ka}{{{\bf K}_\alpha}}
\newcommand{\Gaa}{\Gamma_{\alpha\alpha}}
\newcommand{\Gab}{\Gamma_{\alpha\beta}}
\newcommand{\Gba}{\Gamma_{\beta\alpha}}
\newcommand{\Gbb}{\Gamma_{\beta\beta}}
\begin{document}

\vspace*{1 cm}

\begin{center}
{\LARGE {\bf Mode Coupling in Space and Time Varying

\vspace*{0.25 cm}

Anisotropic Absorbing Plasmas}}

\vspace*{1 cm}

{\Large\sc {\bf {\sc Bijan  Sheikholeslami  Sabzevari}}}

\vspace*{1 cm}

{\large {\bf Institute for Studies in Theoretical Physics and Mathematics\\
P.O. Box 19395-5531, Tehran, Iran}}

\vspace*{7cm}

{\large {\bf Abstract}}
\end{center}
\nl
\begin{quote}
A four dimensional systematic mathematical approach for investigating
propagation and coupling of wave modes in a slowly varying (in all space
directions and time) anisotropic,
absorbing plasma is represented. The formalism is especially useful for energy
considerations of the waves. It is applicable to general cases of mode
conversion in plasmas with general geometries of space-time and magnetic field
configurations.
A simple example of how this formalism can be applied to practical cases is
given.
\end{quote}
\newpage
\section {\bf Introduction}
Almost all investigations about wave coupling and mode conversion in plasmas
were done for cases, where the inhomogeneity of the medium is only in one
direction, i.e. plane stratified plasmas (Stix and Swanson 1983, Swanson
1989, Stix 1992 , Budden 1985). The propagation of electromagnetic waves in
four-dimensional space and time varying plasmas were studied before
(Bernstein \& Friedland 1983, Kravtsov 1969) in the context of geometrical
optics. But in these studies the coupling of modes was not considered and
the mere interest was concentrated an other themes like for example energy
relations and the absorption of independent waves. There exist papers about
four dimensional mode-conversion (Friedland, Goldner \& Kaufman 1987,
Kaufman \& Friedland  1987), but the mathematical proof represented there is
only valid for the very special
cases where the coupling point is not a branch point. The reason is, that the
different dispersion equations combined there result from one unique dispersion
matrix, which is not the case when we consider for example the unified theory
of a class of mode
conversion problems, represented by Cairns \& Lashmore-Davies 1983. Three
dimensional transmission of the fast wave around the ion cyclotron resonance
were considered by Friedland 1990 using reduction techniques.
For recent investigations about combined mode conversion and absorbtion in one
dimensional inhomogeneity see
Cairns et.al.\ 1995 and references therein.
\nl
If we want to study plasma physics more accurate,
the propagation and coupling of electromagnetic waves in media which vary in
more than one dimension have many applications.
The most important application of it is perhaps in
radio-frequency heating and the investigations of plasmas in a tokamak
(Cairns 1991). The plasmas there are considered as plane stratified, but
being precise, we have also to take into account the curvature of the
tokamak. In almost all cases, when wave coupling was under consideration, this
curvature was assumed to be negligible, or at most to give only small and
unimportant effects. But experience in plasma physics
has always shown, that small variations can give huge effects, because
generally plasmas are unstable media. But also in the case of stability,
it has been shown, that small variations can have essential new results, for
example relativistic effects in wave absorptions (Mc Donalds et al 1994).
Another point is, that in the papers about mode resonances, the magnetic field
in the plasma is variable in only one direction. But we know, that in a
okamak, the magnetic field is also toroidal and poloidal. Response
functions for a variable  toroidal magnetic field were calculated by Catto,
Lashmore-Davies and Martin 1993. Another example of the application of wave
coupling with general geometries is the ionosphere, which is usually
considered as
a plane stratified cold plasma (Budden 1985) and is a cost free medium for many
plasma wave observations.
Calculations taking into account the curvature of
the atmosphere could give us new insight in the physics of the ionosphere.
\nl
It is the aim of this
paper to give a systematic derivation of coupled wave equations in slowly
varying plasmas which are three-dimensional inhomogeneous, non stationary,
anisotropic and absorbing.
In order to be self-contained, we include also
a systematic representation of some results
already presented on conferences before (Suchy \& Sabzevari 1992\,a,b) and
refer to these papers for the explicit calculations. Some coupled wave
equations were also represented there, but they must fulfil special conditions
and are not applicable in general, as the derived equations in the present
paper are.
\section { The Eikonal-Maxwell System of Differential Equations}\label{zwei}

In a plasma with time and space dispersion the electric current density,
dielectric displacement and magnetic induction must be written as integral
equations
\ba\label{2.1}
{\bf J}({\bf X})&=&\int d^4\xp{\bf \sigma}^{\mbox{ker}}(\xmx;\x)\cdot
{\bf E}(\xp)
\nonumber\\
{\bf D}({\bf X})&=&\int d^4\xp{\bf \varepsilon}^{\mbox{ker}}(\xmx;\x)\cdot
{\bf E}(\xp)\\
{\bf B}({\bf X})&=&\int d^4\xp{\bf \mu}^{\mbox{ker}}(\xmx;\x)\cdot {\bf H}(\xp)
\nonumber
\ea
where the tensors of electric conductivity, dielectric - and magnetic
permeability
are written as the kernels of the integral equations. They have two kinds of
dependencies on the space-time four-vectors $\x$ and $\xp$. The ones before
the semicolon are fast varying and are due to the collective effects of
the medium, and those after the semicolon are slowly varying and
due to the features of the medium itself, like density etc. In the integration
over time, causality of course must be taken into account, but we redefine it
in the kernels, so that the equations can be written in the above compact
form, where all integrations are from $-\infty \rightarrow +\infty$.
These equation must be solved together with the Maxwell equations

\be\label{2.2}\begin{array}{lcl}
\nabla\times{\bf H}(\x)&=&\dt{\bf D}(\x)+{\bf j}(\x)\\
\nabla\times{\bf E}(\x)&=&-\dt{\bf B}(\x).
\end{array}\ee
\nl
To do this, we first transform the integral equation ~(\ref{2.1}) into
differential equations with the help of geometrical optics (Suchy \&
Sabzevari 1992a, Kravstov 1969), by writing ${\bf j}, {\bf D}$ and ${\bf B}$
in eikonal form
\ba\label{2.3}
{\bf E}(\x)&=&{\bf E}(;\x)\exp\{i\phi(\x)\} \nonumber\\
{\bf H}(\x)&=&{\bf H}(;\x)\exp\{i\phi(\x)\}\\
{\bf j}(\x)&=&{\bf j}(;\x)\exp\{i\phi(\x)\}\nonumber
\ea
where $\phi(\x)$ is the eikonal, which in geometrical optics is a fast
varying function of space and time, and ${\bf E}(;\x), {\bf H}(;\x)$ and
${\bf j}(;\x)$ are the amplitudes, which are slowly varying, because of the
slowness of variation of the medium. The first two equations of ~(\ref{2.3})
can also be considered as the wave entering the plasma, and the last as the
movement of the charged particles in reaction of the waves. On principle,
constant external fields could be added to ~(\ref{2.3}), but
later they give us no new physical results.
\nl
Three different kinds of scales are considered here. First, we suppose
that on each
particle only the particles in the vicinity of that particle in space
and time have an essential influence. This scale we define by
$\Delta _1$, which is small compared to the size of the plasma in space and
time. Hence in ~(\ref{2.1})
\begin{equation}\label{2.4}
\x-\xp=\mbox{O}(\Delta_1)\ll1.
\end{equation}
Then, it is sufficient to integrate only over the vicinity of the point $\x$.
The second scale is due to the slowness of the variation of the medium,
which we denote by a small number $\Delta _2$.  So we can write
\begin{equation}\label{2.5}
\frac{\partial{\bf E}(;\x)}{\partial\x}\ ,\
\frac{\partial{\bf H}(;\x)}{\partial\x}\ ,\
\frac{\partial{\bf G}(;\x)}{\partial\x}\ =\
\mbox{O}(\Delta_2)\ll1
\end{equation}
where ${\bf G}(;\x)$ is an arbitrary function of the medium.
The third scale, which we denote by $\Lambda$, is due to the largeness of
the frequency and wave number, which are defined as the four dimensional
wave vector
\begin{equation}\label{2.6}
{\bf K}(\x)=\frac{\partial\phi(\x)}{\partial\x}.
\end{equation}
In geometrical optics, this can be written as
\begin{equation}\label{2.7}
{\bf K}(\x)=\Lambda\,{\bf G}(;\x) \ \ \ \ \ \ \Lambda\gg 1
\end{equation}
where ${\bf G}(;\x)$ is a slowly varying function
due to the slowness of variation of the medium and is of order O(1). From
~(\ref{2.6}) and ~(\ref{2.7})
\be\label{2.8}
\phi(\x)=\int^{\x}{\bf K}(\xp)\cdot d\xp=\Lambda\int^{\bf X}{\bf G}(;\xp)
\cdot d\xp \approx\mbox{O}(\Lambda).
\ee
Hence, because of ~(\ref{2.5})
\be\label{2.9}
\frac{\partial}{\partial\x}{\bf K}(\x)=\Lambda\dx{\bf G}(;\x)=
\mbox{O}(\Lambda\Delta_2).
\ee
In general, these three scales do not have to be of the same order
\be\label{2.10}
\mbox{O}(\Delta_1)\neq\mbox{O}(\Delta_2)\neq\mbox{O}(\frac{1}{\Lambda}).
\ee
Especially, the variation of the wave vector does not have to be as small along
the ray path, as often is believed and assumed.
\nl
Expanding now $\phi(\xp),{\bf E}(\xp)$ and ${\bf H}(\xp)$ around the point
$\x$, we obtain
\ba\label{2.11}
\lefteqn{{\bf E}(\xp)=e^{i\phi(\x)}e^{i(\xx)\cdot{\bf K}(\x)}\left\{{\bf E}
(;\x)+(\xx)\cdot
\dx{\bf E}(;\x)+\right.} \\ \nonumber
&  & \frac{i}{2}\,\left.\left[(\xx)(\xx):\dx{\bf K}\right]\,{\bf E}(;\x)+
\mbox{O}({\Delta_1}^2\Delta_2)+\mbox{O}(\Lambda\Delta^3{\Delta_2}^2)\right\},
\ea
where we have used dyadic and double scalar products. A similar expression
can be obtained for ${\bf H}(\xp)$. Fourier transforming the fast varying
part of ${\bf \sigma}^{\mbox{ker}}(\xmx;\x),
{\bf \varepsilon}^{\mbox{ker}}(\xmx;\x)$
and ${\bf \mu}^{\mbox{ker}}(\xmx;\x)$ in ~(\ref{2.1}) and using the
features of the
delta function $\delta(\kk)$ and it's derivatives, we can transform
~(\ref{2.1}) into (Suchy \& Sabzevari 1992a, Kravtsov 1969)
\ba\label{2.12}
{\bf j}(\x)&=&e^{i\phi(\x)}[{\bf \sigma}({\bf K};\x)(1-i\ja-
\frac{i}{2}\sp)]\cdot
{\bf E}(;\x) \nonumber\\
{\bf D}(\x)&=&e^{i\phi(\x)}[{\bf \varepsilon}({\bf K};\x)(1-i\ja-
\frac{i}{2}\sp)]\cdot
{\bf E}(;\x) \\
{\bf B}(\x)&=&e^{i\phi(\x)}[{\bf \mu}({\bf K};\x)(1-i\ja-\frac{i}{2}\sp)]\cdot
{\bf H}(;\x) \nonumber
\ea
where $\mbox{O}({\Delta_1}^2\Delta_2)+\mbox{O}(\Lambda\Delta^3{\Delta_2}^2)\}
$ terms have been neglected and the Janus- and Spreading-Operators are
defined correspondingly as
\be\label{2.13}
\ja=\frac{\stackrel{\leftarrow}{\partial}}{\partial{\bf K}}\cdot
\frac{\stackrel{\rightarrow}{\partial}}{\partial{\bf X}}
\ \ \ \mbox{and}\ \ \  \sp=\frac{\stackrel{\leftarrow}{\partial^2}}{
\partial{\bf K}\partial{\bf K}}:
\frac{\partial}{\partial{\bf X}}{\bf K}.
\ee
\nl
To combine them with the Maxwell equations, we first have to calculate the
time derivatives of ${\bf D}(\x)$ and ${\bf B}(\x)$. To do this, we first
write ${\bf D}(\x)$ (and correspondingly ${\bf B}(\x)$) in the form
(Suchy \& Sabzevari 1992\,a, Kravtsov 1969)
\be\label{2.14}
{\bf D}={\bf D}({\bf K};\x)\exp\{i\phi(\x)\}
\ee
with
\be\label{2.15}
{\bf D}({\bf K};\x)=[{\bf \varepsilon}({\bf K};\x)(1-i\ja-\frac{i}{2}\sp)]\cdot
{\bf E}(;\x)+
\mbox{O}({\Delta_1}^2\Delta_2)+\mbox{O}(\Lambda{\Delta_1}^3{\Delta_2}^2)
\ee
{}From ~(\ref{2.7}), we can write
\be\label{2.16a}
\frac{\partial}{\partial{\bf K}}=
\frac{1}{\Lambda}
\frac{\partial}{\partial{\bf G}}=\mbox{O}(\frac{1}{\Lambda})
\ee
\be\label{2.16b}
\frac{\partial^2}{\partial{\bf K}\partial{\bf K}}=
\frac{1}{\Lambda^2}
\frac{\partial^2}{\partial{\bf G}\partial{\bf G}}=
\mbox{O}(\frac{1}{\Lambda^2}),
\ee
\nl
and from the last two equations follows
\be\label{2.17}
\frac{\partial}{\partial t}{\bf D}({\bf K};\x)=
\frac{\partial}{\partial t}[{\bf \varepsilon}({\bf K};\x)\cdot
{\bf E}(;\x)]+\mbox{O}(\frac{\Delta_2}{\Lambda}).
\ee
In all equations until now, we kept terms of order O($\Delta_1\Delta_2$),
i.e.\ terms which are products of two small numbers. The last term in
~(\ref{2.17}) is also the product of two small numbers. But we can neglect
it compared with the last one,
since $\mbox{O}(1/\Lambda)\ll\mbox{O}(\Delta_1)$. This follows from
the fact that in geometrical optics the scale of the wavelenght and period
can be considered as much smaller than the scale of the distance and time,
on which
different parts of the medium have an essential influence on each other.
Using the result ~(\ref{2.17}) and corresponding expressions for the magnetic
induction, we finally obtain
\ba\label{2.18}
\dt{\bf D}(\x)=\exp\{i\phi(\x)\}\ep\left\{-i\omega(\x)\ep\right.
 & \\  &  -\left[\omega(\x)\ep(\ja+\frac{1}{2}\sp)+
\left[\dt\ep\right]_{\bf K}\right\}\cdot{\bf E}(;\x) \nonumber
\ea
and a corresponding expression for $\partial{\bf B}(\x)/\partial t$, where
${\bf \varepsilon}$ is interchanged with ${\bf \mu}$ and ${\bf E}(;\x)$ with
${\bf H}(;\x)$. The time derivatives of ${\bf D}(\x)$ and ${\bf B}(\x)$
together with ${\bf j}(\x)$ from ~(\ref{2.12}) can now be combined
with the Maxwell equations ~(\ref{2.2}) to give a system a partial
differential equation for
the components of the six dimensional  electromagnetic field amplitude
\be\label{2.19}
{\bf f}(;\x)=\left[\begin{array}{c} {\bf E}(;\x)\\{\bf H}(;\x)\end{array}
\right].
\ee
Since we have different modes propagating in the medium, we first expand
the electromagnetic field into these modes
\be\label{2.20}
{\bf f}(\x)=\sum_\alpha\epha\,{\bf f}_\alpha(;\x)
\ee
where $\phi_\alpha(\x)$ is the eikonal and ${\bf f}_\alpha(;\x)$
the electromagnetic amplitude for each
mode and each mode has a separate wave vector
\be\label{2.21}
{\bf K}_\alpha(\x)=\frac{\partial\phi_\alpha(\x)}{\partial \x}.
\ee
Then the system of partial differential equations for the electromagnetic
amplitudes of the modes can be represented as
\be\label{2.22}
\sum_\alpha \exp\{i\phi_\alpha(\x)\}
\left\{\Ma(i+\jaa+\frac{1}{2}\spa)-\left[\dt{\bf C}(\ka;\x)
\right]_{\bf K_\alpha}
\right\}\cdot{\bf f}_\alpha(;\x)=0
\ee
which we call the Eikonal-Maxwell system (Suchy \& Sabzevari 1992a), with
the Maxwell tensor
\be\label{2.23}
\Ma= \left(
\begin{array}{cc}
\omega_{\scriptsize{\alpha}}({\bf k};\x)\,{\bf \varepsilon}(\ka;\x)+i\,{\bf
\sigma}(\ka;\x) &
{\bf k}\times {\bf I}\\
    &     \\
-{\bf k}\times {\bf I} & \omega_{\scriptsize{\alpha}}({\bf k};\x)\,{\bf \mu}
(\ka;\x)
\end{array}
\right),
\ee
where ${\bf I}$ is the unit tensor and ${\bf k}$ is the three
dimensional spatial part of the wave vector ~(\ref{2.21}) and
$\omega_{\scriptsize{\alpha}}$
the frequency
\footnote{In $\ka$, it is sufficent to denote the
index $\alpha$ to one component only, because of the dispersion relation
~(\ref{3.4}).}.
${\bf C}(\ka;\x)$ is the material tensor
\be\label{2.24}
{\bf C}(\ka;\x)=\left(
\begin{array}{cc}
\epa & 0\\
&   \\
0 & {\bf \mu}(\ka;\x) \end{array}
\right)_.\ee
$\jaa$ and $\spa$ are the operators ~(\ref{2.13}) corresponding to each mode.
The  derivatives to each mode are defined as
\be\label{2.25}
\left.\frac{\partial}{\partial{\bf K}_\alpha} \Ma=
\frac{\partial}{\partial{\bf K}} {\bf M}({\bf K};\x)\right|_{{\bf K}=\ka}
\ee
\section{The eigenvalue spectrum of the modes and polarization vectors}
\label{drei}
It can be shown (Suchy \& Sbzevari 1992a) that the Maxwell-Tensor
~(\ref{2.23}) can be written as the sum of two parts
\be\label{3.1a}
\M=\MT-\lambda\,\ML
\ee
where we habe devided the wave-vector into a longitudinal and transversal part
\be\label{3.1}
\bf K=\kt+\lambda\,{\bf g}^\lambda
\ee
where ${\bf g}^\lambda$ is the axis in the direction of the longitudinal part.
For example, $\lambda$  can be chosen as minus the frequency $\omega$.
Then ${\bf K}_T$ is the spatial wave vector ${\bf k}$ and we call these
frequency
modes. For the case of stratified (not necessarily plane) media $\lambda$
can be chosen as the component of ${\bf k}$ perpendicular to the
stratification surfaces. For a thorough discussion of various possibilities
see Suchy \& Sabzevari 1992a. It should be noted, that using the method
represented here, restricts the dispersion of ${\bf \epsilon}$, ${\bf \sigma}$
and ${\bf \mu}$ to the three components of $\kt$.
\nl
The form ~(\ref{3.1a}) of the Maxwell tensor is called a generalized
characteristic matrix. We can define right and left eigenvectors
\be\label{3.3}\begin{array}{lcl}
\Ma\cdot\ga & = & [\MT-\la\,\ML]\cdot\ga=0\\      &  &  \\
\gba\cdot\Ma & = & \gba\cdot[\MT-\la\,\ML]=0.
\end{array}\ee
Equations ~(\ref{3.3}) have nontrivial solutions only when
\be\label{3.4}
\mbox{Det}\,\Ma=0.
\ee
This gives us the dispersion relation.
We choose the $\ga$ as the polarization vectors. The advantage of this
choose is, that the right eigenvectors are parallel to the electromagnetic
field in a homogeneous medium, as can be seen immediately from ~(\ref{2.22}).
We can then devide the space-time into small cells and in each cell the
medium can be considered as homogeneous. When we go from one cell to the
other, the $\ga$ will change direction and size. So the first equation of
~(\ref{3.3})
gives us the direction (but not the amount) of the wave field at each
space-time point. The left eigenvectors have in general
no real physical meaning. They are only useful for further calculations.
But they can be considered as waves in a concomitant space (Suchy \& Altman
1975).
\nl
It is easy to prove the following biorthogornality relation
\be\label{3.5}
\gba\cdot{\bf M}_L\cdot\gb=\delta_{\alpha\beta}\,\gba\cdot{\bf M}_L\cdot\ga.
\ee
The right and left eigenvectors are identical when the Maxwell tensor is
equal to it's own transpose. In the case of a hermitean Maxwell tensor, the
left eigenvectors become the complex conjugate of the right eigenvectors.
\nl
Since the right (left) eigenvectors can be defined as an arbitrary linear
combination of the columns (rows) of the adjoint of $\Ma$, we can define
\nl
\be\label{3.6}\begin{array}{lcl}
\ga & = & \frac{\mbox{Adj}\,\Ma\cdot{\bf c}}{\sqrt{{\bf \bar{c}}
\cdot\mbox{Adj}\Ma\cdot{\bf c}}}\\ & & \\
\gba & = & \frac{\bar{\bf c}\cdot\mbox{Adj}\,\Ma}{\sqrt{\bar{\bf c}\cdot
\mbox{Adj}\Ma\cdot
{\bf c}}} \end{array}
\ee
\nl
where $\bar{\bf c}$ and ${\bf c}$ are arbitrary constant column and row
vectors.
An expression for the biorthogonality relation ~(\ref{3.5})
can be obtained (Suchy \& Sabzevari 1992b)
\be\label{3.6c}
\gba\cdot{\bf M}_L\cdot\ga=\Theta\prod_{\gamma\neq\alpha}
(\lambda_\alpha-\lambda_\gamma).
\ee
The factor $\Theta=\Theta({\bf K}_T;\x)$ occures because we have a
generalized characteristic
matrix, i.e.\ the matrix ~(\ref{3.1a}). With this expression, it is possible
to normalize the
polarization vectors in a suitable form
\be\label{3.7}\begin{array}{lll}
\fa= & \frac{\ga}{\sqrt{\gba\cdot{\bf M}_L\cdot\ga}}= &
\frac{\ga}{\sqrt{\Theta\prod_{\gamma\neq\alpha}(\lambda_\alpha-
\lambda_\gamma)}}\\       & & \\
\fba= & \frac{\gba}{\sqrt{\gba\cdot{\bf M}_L\cdot\ga}}= &
\frac{\gba}{\sqrt{\Theta\prod_{\gamma\neq\alpha}(\lambda_\alpha-
\lambda_\gamma)}}\end{array}
\ee
and from the biorthogonality relations ~(\ref{3.5}) follow biorthonormality
relations
\be\label{3.8}
\left\{
\begin{array}{ll}
\fba\cdot{\bf M}_L\cdot\fa=1 & \mbox{always}\\
    &    \\
\fba\cdot{\bf M}_L\cdot\fb=0 & \lambda_{\alpha}\neq\lambda_{\beta}.
\end{array}\right.\ee
We define now  the complete electromagnetic wave amplitude for each mode
by
\be\label{3.9}
{\bf f}_\alpha=a_\alpha(;\x)\,\fa(\ka;\x)
\ee
where $a_\alpha$ is the scalar amplitude of the wave. To write
the modes in the form of the last equation is of great advantage.
The reason is,
that $\bar{\bf f}_\alpha\cdot{\bf M}_L\cdot{\bf f}_\alpha$ is equal to the
energy of the wave
propagating through the medium (Suchy \& Sabzevari 1992a) when collisions are
neglected, i.e. hermitean Maxwell tensor.
Hence, because of the first of the biorthonormality rations in ~(\ref{3.8}),
$|a_\alpha|^2$ is identical to the wave energy of the mode $\alpha$.
In the case of non-hermitean
Maxwell tensor, this last quantity is proportional to the wave energy at each
space-time point, hence $|a_\alpha|^2/|a_\beta|^2$ gives us directly
transmission-,
reflection- or absortbtion- coefficients of different modes.
\section{Mode coupling point}\label{vier}
Since the $\lambda$'s are functions of space and time, there are some points
$\x_0$, where $\lambda_\alpha(\kt(\x_0);\x_0)=\lambda_\beta(\kt(\x_0);\x_0)$.
Then from ~(\ref{3.3}) and ~(\ref{3.7}) the polarization vectors $\fa$ and
$\fb$ become parallel when the rank of the matrix $\Ma\,(=\Mb)$ is equal to 5.
This is because if the rank of a matrix is equal to the dimension minus
the number of linearly independent eigencolumns of the matrix, then we have
only one linearly independent $\fa\,(=\fb)$. Otherwise, there would be more
than one linearly independent eigencolumns for each mode. For example, if the
rank would be 4, then we would have two linearly independent eigencolumns
for each $\alpha$ and $\beta$, i.\,e.\
${\bf f}_{\alpha1}\,(={\bf f}_{\beta1})$ and
${\bf f}_{\alpha2}\,(={\bf f}_{\beta2})$. Hence ${\bf f}_{\alpha1}$ would be
linearly independent of ${\bf f}_{\beta2}$. In this case, at the point
$\x_0$, we can find two vectors $\fa$ and $\fb$, corresponding to different
modes, which are linearly
independent, i.e.\ not parallel. No mode coupling would occure in this case.
The two modes would pass by without any influence on each other (Budden 1985,
chs.\,17.6 and 17.7, Budden \& Smith 1974). Hence the exact definition of
a coupling point is
\be\label{4.1}
\mbox{Coupling point}\ \ \Leftrightarrow\ \ \left\{
\lambda_\alpha=\lambda_\beta\ \ \mbox{for}\ \ \x=\x_0\ \ \mbox{and}\ \
\mbox{Rank}\,\Ma=\mbox{Rank}\,\Mb=5\right\}.\ee
It should be noted, that at the coupling points the scalar amplitudes
$a_\alpha$ and $a_\beta$ are generally not equal. Physically, coupling does
not occure at one mathematical point. Actually there is a small area around
the coupling point, where coupling takes place. This  coupling area
can be defined as
\be\label{4.2}
\mbox{Coupling area}\ \ \Leftrightarrow\ \ \left\{|\x-\x_0|<\epsilon\
\Longleftrightarrow\ |\lambda_\alpha-\lambda_\beta|<
\delta\ \mid\ \epsilon,\delta\ll1,\ \mbox{and}\ \ \lambda_\alpha(\x_0)=
\lambda_\beta(\x_0)\right\}.
\ee
It is possible to estimate the size of $\epsilon$ and $\delta$
\footnote{work is in progress and will be represented in the framework of an
arbitrary number of waves coupling to each other in a separate paper}.
When two waves reach the border of this area, the coupling is weak. The
more they get into the area, the stronger the coupling will be. At the
coupling point, the coupling is maximum. Usually, a singularity is to be
expected there (for the example of a stratified medium see Sabzevari 1992,
1993, Budden 1985, ch.\,16.3, Budden 1972)
\nl
{}From ~(\ref{3.7}) and ~(\ref{3.8}) follow the biorthonormality relations
\be\label{4.3}
\begin{array}{lcl}
\fba\cdot{\bf M}_L\cdot\fb=1 & \mbox{for} & \x=\x_0\\
   &   &  \\
\fba\cdot{\bf M}_L\cdot\fb=0 & \mbox{for} & \x\neq\x_0.
\end{array}\ee
As we can see from ~(\ref{4.3}), the point $\x_0$ is a
``jump singularity'', because the scalar product is discontinuos there.
\section{Transport equations for the amplitude}\label{fuenf}
Since the polarization vectors $\fa$ can be calculated by the local
eigenvector equation ~(\ref{3.3}) at each space-time point $\x$, we need
another equation to calculate the scalar amplitudes $a_\alpha$ along the
rays to obtain the complete wavefield ~(\ref{3.9}) at each space-time
point. Putting ~(\ref{3.9}) into the system ~(\ref{2.22}), a system of
partial differential equations for the amplitudes can be obtained (for the
technical details of the calculations see Suchy \& Sabzevari 1992a)
\be\label{5.1}
\sum_\gamma\ephg\left[{\bf W}_{\rho\gamma}\cdot
\frac{\partial a_\gamma}{\partial\x}+\Gamma_{\rho\gamma}a_\gamma
\right]=0.\ee
The transport vectors ${\bf W}_{\rho\gamma}$ give the change of the amplitude
in the direction of this vector, as can be seen from the last equation itself.
They have
the form
\be\label{5.2}
{\bf W}_{\rho\gamma}=\delta_{\rho\gamma}\frac{d\x}{d\tau_\gamma}+
(\lambda_\rho-\lambda_\gamma)\,\fbr\cdot{\bf M}_L\cdot(\fg\,\dkt),\ee
and the coefficients of the system are
\be\label{5.3}
\Gamma_{\rho\gamma}=
\delta_{\rho\gamma}
\frac{\stackrel{\rightarrow}{\partial}_\gamma{\bf D}_\gamma}{2}+
(\lambda_\rho-\lambda_\gamma)\,\fbr\cdot{\bf M}_L\cdot
\left(
\fg\,\frac{\stackrel{\leftarrow}{d}}{d\kt d\kt}\,:\,
\frac{\partial}{\partial\x_T}\kt
\right)
+{\hat{\Gamma}}_{\rho\gamma}  \ee
with
\ba\label{5.4}
\lefteqn{
{\hat{\Gamma}}_{\rho\gamma}\,=\,\fbr\cdot\left[{\bf M}_L\cdot
\frac{d\fg}{d\tau_\gamma}\,+\,\left(\frac{d {\bf M}_L}{d\tau_\gamma}\right)_\x
\cdot\fg\right.} \nonumber \\
 & &  -{\bf M}_\gamma\,\dkt\,:\,
\left(\frac{\partial}{\partial\x_T}\fg\right)_\kt+
\left.\left(\frac{\partial{\bf C}}{\partial t}\right)_{{\bf K}_\gamma}
\cdot\fg\right].
\ea
In these relations $\x=\x_T+{\bf g}_\lambda\x_L$ with ${\bf g}_\lambda
\cdot{\bf g}^\lambda=1$ (compare  with ~(\ref{3.1})). $\tau_\gamma$ is the
parametrization of the ray $\gamma$ and is defined via the Hamiltonian
equations (Suchy \& Sabzevari 1992a, Berstein \& Friedland
1983)
\nl
\be\label{5.5}\begin{array}{lcl}
\frac{d\x}{d\tau_\gamma} & = & \frac{\partial D_\gamma({\bf K};\x)}{\partial
{\bf K}}\\     & & \\
\frac{d{\bf K}}{d\tau_\gamma} & = & -\left[\frac{\partial
D_\gamma({\bf K};\x)}{\partial\x}
\right]_{\bf K},
\end{array}\ee
\nl
where $D_\gamma$ is defined as
\be\label{5.6a}
\mbox{Det}\,\M= -\Theta(\kt;\x)\prod_\rho D_\gamma({\bf K};\x)
\ee with
\be\label{5.6b}
D_\gamma({\bf K};\x)=\lambda-\lambda_\gamma(\kt;\x).
\ee
\nl
In the case where all the modes are propagating independently,
i.e.\,when there is no mode
coupling, there are no points where the ray paths coincide and the paths are
independent from each other. In this case the functions
\be\label{5.7}
\phi_\gamma(\x)=\int_{\mbox{\scriptsize ray path}}^\x
d\x^\prime\cdot{\bf K}_\gamma(
\x^\prime)\ee
are independent from each other. The second terms in ~(\ref{5.2}) and
~(\ref{5.3})
become identical to zero and the Kronecker symbol becomes one. From
~(\ref{5.1}) follows then an
ordinary differential equation for the scalar amplitude of each mode $\alpha$
along it's ray path
\be\label{5.8}
\frac{d a_\alpha}{d \tau_\alpha}+\Gamma_{\alpha\alpha}\,a_\alpha=0
\ee
with
\be\label{5.9}
\Gamma_{\alpha\alpha}\,=\,\frac{\stackrel{\rightarrow}{\partial_\alpha}
D_\alpha}{2}\,+\,{\hat{\Gamma}}_{\alpha\alpha}.
\ee
The solution is
\be\label{5.10}
a_\alpha(\tau_\alpha)\,=\,\mbox{const}\,\exp{\left\{-\int^{\tau_\alpha}
d\tau_\alpha^\prime\,\Gaa\right\}}\ee
which gives us the amplitude along it's path. For more discussions about the
solutions of independent modes see Suchy \& Sabzevari 1992a, sec.\,7,
Sabzevari 1992, 1993.
\nl
Now the advantage of the representation ~(\ref{3.9}) (except that it is
four-dimensional in space and time) becomes
more transparent, since with the last equations, we can calculate the change
of the scalar amplitude, and hence the energy directly, without bothering
about polarization vectors
\section{Coupled wave equations}\label{sechs}
In this section, we explicitly derive coupled ordinary differential equations
for the amplitudes of coupled waves near their coupling point (except the
coupling point itself, because there is a jump singularity, but we can
approach the coupling point infinitly), i.\,e.\
where  $\lambda_\alpha\approx
\lambda_\beta$.
\nl
In the case of the coupling of the two modes $\alpha$ and $\beta$, we can see
from ~(\ref{5.7}) that $\phi_\alpha$ and $\phi_\beta$ can not be considered as
independent anymore, because their ray path coincide in $\x$- and
${\bf K}$-space at the coupling point
$\x_0$. Since an arbitrary constant phase can be added to the eikonals, we can
write near the coupling point in ${\bf K}$-space
\be\label{6.1}
\phi_\alpha(\x)=\int_{\x_0}^\x d\x^\prime\cdot{\bf K}_\alpha(\x^\prime)
\approx\int_{\x_0}^\x d\x^\prime\cdot{\bf K}_\beta(\x^\prime)=\phi_\beta(\x).
\ee
As far as only the two modes couple and are decoupled of the other modes,
$\phi_\alpha$ and $\phi_\beta$ are independent of the eikonals corresponding to
the other modes. From ~(\ref{5.1}) follows then
\nl
\be\label{6.2} \begin{array}{ll}
{\bf W}_{\alpha\alpha}\cdot \frac{\partial a_\alpha}{\partial\x}+{\bf W}_{
\alpha\beta}\cdot\frac{\partial a_\beta}{\partial\x}+\Gaa a_\alpha+\Gab a_\beta
 & =0  \\ &  \\
{\bf W}_{\beta\alpha}\cdot \frac{\partial a_\alpha}{\partial\x}+{\bf W}_{
\beta\beta}\cdot\frac{\partial a_\beta}{\partial\x}+\Gba a_\alpha+\Gbb a_\beta
 & =0.\end{array}\ee
 \nl
 For the transport vectors follows from ~(\ref{5.2})
 \be\label{6.3}
 {\bf W}_{\alpha\beta}=(\lambda_\alpha-\lambda_\beta)\,\fba\cdot{\bf M}_L\cdot
 \left(\fb\dkt\right).\ee
 It may seem at the first moment, that this term is small, because
 $\lambda_\alpha\approx\lambda_\beta$.
 But this assumption is invalid, as we will see soon. To calculate the
 transport vector, we use ~(\ref{3.6c}) to write the equations ~(\ref{3.7}) as
 \be\label{6.4}\begin{array}{lcl}
 \fb & = & (\lambda_\beta-\lambda_\alpha)^{-\frac{1}{2}}\,
 \frac{\gb}{\sqrt{\Theta\prod_{\gamma
 \neq\alpha,\beta}\left(\lambda_\beta-\lambda_\gamma\right)}}\\
    &    &    \\
 \fba & = & (\lambda_\alpha-\lambda_\beta)^{-\frac{1}{2}}\,
 \frac{\gba}{\sqrt{\Theta\prod_{\gamma
 \neq\alpha,\beta}\left(\lambda_\alpha-\lambda_\gamma\right)}}
 \end{array}\ee
 {} From ~(\ref{3.5}) and ~(\ref{3.6c}) it is clear that
 \be\label{6.5}
 \gba\cdot{\bf M}_L\cdot\gb=0\ \ \ \mbox{for all}\ \ \lambda_\alpha\ \mbox{and}
 \ \lambda_\beta
 \ee
 We differentiat $\fb$ to $\kt$ and multiply it from the right with
 $\fba\cdot\ML$ to obtain with ~(\ref{6.5}) for ~(\ref{6.3})
 \be\label{6.6}
 {\bf W}_{\alpha\beta}\,=\,i\,\frac{\gba\cdot{\bf M}_L\cdot
 \left(\gb\dkt\right)}{\Theta
 \sqrt{\prod_{\gamma\neq\alpha,\beta}\,(\lambda_\alpha-\lambda_\gamma)
 (\lambda_\beta-\lambda_\gamma)}}.
 \ee
Since $\gb$ is an arbitrary linear combination of the columns of Adj${\bf
M}({\bf K}_\beta;\x)$, the elements of this column are polynomials of
$\lambda_\beta$. Hence
\be\label{6.7}
\gb\dkt\,=\,\left(\frac{\partial}{\partial\lambda_\beta}\gb\right)
\left(\frac{\partial\lambda_\beta}{\partial\kt}\right)\,+\,\left(\gb
\frac{\stackrel{\leftarrow}{\partial}}{\partial\kt}\right)_{\lambda_\beta}.
\ee
It can be shown (see appendix) that near the coupling point in general
\be\label{6.8}
\frac{\partial\lambda_\beta}{\partial\kt}\ \propto\ \frac{1}{\lambda_\alpha-
\lambda_\beta}.
\ee
This relation is not only valid for the derivatives to $\kt$, but for the
derivatives
of all kind of variables. The coefficients of the polynomials in
$\gb$ can in general be assumed as analytical, since they depend only on the
elements of the matrices $\MT$ and $\ML$, which depend only on the
${\bf \epsilon}, {\bf \mu}, {\bf \sigma}, \kt$ and $\x$.
Therefore the first
term in ~(\ref{6.7}) in general is growing relativ to the second as we
approach the
coupling point. Hence, we can write
\be\label{6.9}
\gb\dkt\approx\left(\frac{\partial}{\partial\lambda_\beta}\gb\right)
\left(\frac{\partial\lambda_\beta}{\partial\kt}\right).
\ee
Now from the first Hamiltonian equation ~(\ref{5.5}) and the definition
of $D_\beta$ from ~(\ref{5.6b}) follows
\be\label{6.10}
\frac{d\x}{d\tau_\beta}=-\frac{\partial\lambda_\beta}{\partial\kt}
+{\bf g}_\lambda
\ee
Near the coupling point ~(\ref{6.8}) can be used to write ~(\ref{6.10}) as
\be\label{6.11}
\frac{d\x}{d\tau_\beta}\approx-\frac{\partial\lambda_\beta}{\partial\kt}.
\ee
Putting this in ~(\ref{6.9}), it follows from ~(\ref{6.6})
\be\label{6.12}\begin{array}{lr}
{\bf W}_{\alpha\beta}\cdot\frac{\partial a_\beta}{\partial\x}=
\Upsilon_1\,\frac{d a_\beta}{d\tau_\beta} & \ \ \ \ \ {\bf W}_{\alpha\alpha}
\cdot
\frac{\partial}{\partial\x}=\frac{d}{d\tau_\alpha}\\
     &    \\
{\bf W}_{\beta\alpha}\cdot\frac{\partial a_\alpha}{\partial\x}=
\Upsilon_2\,\frac{d a_\alpha}{d\tau_\alpha} & \ \ \ \ \ {\bf W}_{\beta\beta}
\cdot
\frac{\partial}{\partial\x}=\frac{d}{d\tau_\beta}
\end{array}\ee
where
\nl
\be\label{6.13}\begin{array}{lclcl}
\Upsilon_1 & = &
\frac{(\lambda_\alpha-\lambda_\beta)\gba\cdot\Ml\cdot\left(
\frac{\partial}{\partial
\lambda_\beta}\gb\right)}{\sqrt{\gba\cdot{\bf M}_L\cdot\ga}
\sqrt{\gbb\cdot{\bf M}_L\cdot\gb}} & = &
-i\frac{\gba\cdot\Ml\cdot\left(\frac{\partial}{\partial
\lambda_\beta}\gb\right)}{\Theta\,\sqrt{\prod_{\gamma\neq\alpha,\beta}
\left(\lambda_\alpha-\lambda_\gamma\right)\left(\lambda_\beta-
\lambda_\gamma\right)}}\\ & & & & \\
\Upsilon_2 & = &
\frac{-(\lambda_\alpha-\lambda_\beta)\gbb\cdot\Ml\cdot\left(
\frac{\partial}{\partial
\lambda_\alpha}\ga\right)}{\sqrt{\gba\cdot{\bf M}_L\cdot\ga}
\sqrt{\gbb\cdot{\bf M}_L\cdot\gb}} & = &
+i\frac{\gbb\cdot\Ml\cdot\left(\frac{\partial}{\partial
\lambda_\alpha}\ga\right)}{\Theta\,\sqrt{\prod_{\gamma\neq\alpha,\beta}
\left(\lambda_\alpha-\lambda_\gamma\right)\left(\lambda_\beta-
\lambda_\gamma\right)}}
\end{array}
\ee
\nl
for $\lambda_\alpha\approx\lambda_\beta$ and $\lambda_\alpha>\lambda_\beta$
(when $\lambda_\alpha,<\lambda_\beta$, the sign of $\Upsilon_1$ and
$\Upsilon_2$ would be interchanged). When only the two modes $\alpha$
and $\beta$ couple, whilst the others are decoupled, $\Upsilon_1$ and
$\Upsilon_2$ are finite. In contrast, by the same methods represented here,
it can be shown that the $\Gab$, $\Gba$, $\Gaa$ and $\Gbb$ are growing
to infinity as we approach the coupling point. This is due to the second
derivative of the second term on the right hand side of ~(\ref{5.3}).
{}From ~(\ref{6.8}) this
second derivative is of order O($\lambda_\alpha-\lambda_\beta)^{-2}$ and
the singularity can not be removed when multiplied by $(\lambda_\alpha-
\lambda_\beta)$.
\nl
We are now ready to construct the coupled equations. Near the coupling
points the paths almost coincide and we can write $d\tau_\alpha\approx
d\tau_\beta=d\tau$. Putting ~(\ref{6.12}) in ~(\ref{6.2}), we obtain after
some subtractions and additions the system
\be\label{6.14}
\frac{d}{d\tau}{\bf a}\,-\,{\bf \Gamma}\cdot{\bf a}=0
\ee
where
\be\label{6.15}
{\bf a}\,=\,\left(\begin{array}{c}a_\alpha\\a_\beta\end{array}\right)
\ \ \ \ \  \ \
{\bf \Gamma}=\left(\begin{array}{cc} \Gamma_{11} & \Gamma_{12}\\
\Gamma_{21} & \Gamma_{22} \end{array}\right)\ee
with
\be\label{6.16}\begin{array}{ll}
\Gamma_{11}=\frac{\Upsilon_1\Gamma_{\beta\alpha}-\Gamma_{\alpha\alpha}}{1-
\Upsilon_1\Upsilon_2} &
\ \ \ \ \Gamma_{12}=\frac{\Upsilon_1\Gamma_{\beta\beta}-\Gamma_{\alpha\beta}}{
1-\Upsilon_1\Upsilon_2} \\ & \\
\Gamma_{21}=\frac{\Upsilon_2\Gamma_{\alpha\alpha}-\Gamma_{\beta\alpha}}{1-
\Upsilon_1\Upsilon_2} &
\ \ \ \ \Gamma_{22}=\frac{\Upsilon_2\Gamma_{\alpha\beta}-\Gamma_{\beta\beta}}{
1-\Upsilon_1\Upsilon_2}.
\end{array}\ee
The solution of ~(\ref{6.14}) is
\be\label{6.17}
{\bf a}=\exp\left\{\int^\tau{\bf \Gamma}\,d\tau^\prime\right\}\cdot{\bf a}_0
\ee
where ${\bf a}_0$ is an arbitrary constant column.
\nl
For a stratified medium, a different formalism than here, which was constructed
for a coupling of an arbitrary number of coupled waves,  also resulted in
coupled wave equations with singular coefficients (Sabzevari 1992, 1993).
But the form of the equation were of a kind, that the singularities first
had to be removed in order to solve the problem (see also Budden 1985,
ch. 16 and 17, Friedland 1985). This is not the case here in general.
An example is given below. We believe,
that the origin of these singularities, not only for the present case, but for
all other cases studied before, could stem from the linearization of the
coupling. Perhaps, these singularities would not appear, if the problem
would be reconsidered under the aspect of nonlinearity (Tskhakaya 1996).
These topics could
be the basis for further investigations. The best method to investigate
mode-conversion by the system of equations ~(\ref{6.14}) is to use S-matrix
methods. Using
this method, there is no need to bother about the singularities at the
coupling point. For an example see below.
\section{The example of a plane-stratified plasma}
To give a simple example of how to handle with the above formulas in practical
cases and also to have a cheque of the correctness of the formalism, we
consider the well known case of a plane stratified isotropic collisionless
cold plasma with an arbitrary number of ion components. The magnetic field is
in the $z$-direction normal to the stratification surfaces. The wave is in
the $x-z$-plane and $\theta$ the angle between the $k$-vector and the
$z$-direction.
In this case,
the Maxwell tensor has the simple form (Swanson 1989, ch. 2.1.2)
\be\label{7.1}\left[
\begin{array}{cccccc}
\omega\epsilon_0 P & 0 & 0 & 0 & -k\cos\theta & 0 \\
0 & \omega\epsilon_0 P & 0 & k\cos\theta & 0 & -k\sin\theta\\
0 & 0 & \omega\epsilon_0 P & 0 & k\sin\theta & 0 \\
0 & k\cos\theta & 0 & \omega\mu_0 & 0 & 0 \\
-k\cos\theta & 0 & k\sin\theta & 0 & \omega\mu_0 & 0 \\
0 & -k\sin\theta & 0 & 0 & 0 & \omega\mu_0
\end{array} \right]\ee
where
\be\label{7.1a}
P=1-\sum_j \frac{\omega_{pj}^2}{\omega^2}.
\ee
We consider monochromatic wave-guide modes, i.e.\ the component of the wave
vector in the $z$-direction. Hence $\lambda=k_z=k\cos\theta$ and from Snell's
law ${\bf K}_T=k_x=k\sin\theta=(\omega/c_0)\sin\theta_0=$const., where $c_0$
is the speed of light in vacuum and $\theta_0$ the
angle, the wave makes with the $z$-direction before entering the plasma, i.e.\
in the vacuum. Then, we have
\be\label{7.2}
\Ml=\left[ \begin{array}{cccccc}
0 & 0 & 0 & 0 & 1 & 0 \\
0 & 0 & 0 & -1& 0 & 0 \\
0 & 0 & 0 & 0 & 0 & 0 \\
0 & -1& 0 & 0 & 0 & 0 \\
1 & 0 & 0 & 0 & 0 & 0 \\
0 & 0 & 0 & 0 & 0 & 0
\end{array}\right].\ee
The coefficient ~(\ref{5.3}) reduce to
\be\label{7.2a}
\Gamma_{\rho\gamma}\,=\,\fbr\cdot\Ml\cdot\left(\frac{d}{dz}\,\fg\right).
\ee
To derive the right and left eigenvectors ${\bf g}$ and $\bar{\bf g}$, it
suffices first to calculate the second row and column of Adj\,${\bf M}$
respectivly. We define then
\be\label{7.3}
\bar{c}\,=\,(0,1,0,0,0,0)\ \ \ \ \ \mbox{and}\ \ \ \ \ c\,=\,\left[
\begin{array}{c}
0\\1\\0\\0\\0\\0\end{array}\right]\ee
and obtain  from  ~(\ref{3.6})
\be\label{7.4}
\ga\,=\,\gba\,=\,\frac{\omega^3}{c_0^2}\sqrt{\frac{\mu_0}{\omega}}
\left[ \begin{array}{l}
0 \\ \ \\ (P(P-n_\alpha^2))^{\frac{1}{2}}\\ \ \\
0 \\ \ \\ \sqrt{\frac{\epsilon_0}{\mu_0}}(P(P-n_\alpha^2))^\frac{1}{2}n_\alpha
\cos\theta\\ \ \\
0 \\ \ \\ \sqrt{\frac{\epsilon_0}{\mu_0}}(P(P-n_\alpha^2))^\frac{1}{2}n_\alpha
\sin\theta
\end{array} \right] \ee
where $n_\alpha=(c_0/\omega){k}_\alpha$ is the refractive index.
{}From ~(\ref{3.7}) we obtain
\be\label{7.6}
\fa\,=\,\fba\,=\,\sqrt{\frac{c_0\mu_0}{2}}
\left[ \begin{array}{l}
0 \\ \ \\ (n_\alpha\cos\theta)^{\frac{-1}{2}}\\ \ \\
0 \\ \ \\ \sqrt{\frac{\epsilon_0}{\mu_0}}(n_\alpha\cos\theta)^\frac{1}{2}\\ \
\\
0 \\ \ \\ \sqrt{\frac{\epsilon_0}{\mu_0}}n_\alpha^\frac{1}{2}\sin\theta(\cos
\theta)^\frac{-1}{2}
\end{array} \right]. \ee
Using ~(\ref{7.4}), the equations ~(\ref{6.13}) can be calculated
\be\label{7.7}
\Upsilon_{\stackrel{\scriptsize 1}{\scriptsize 2}}\,=\,\pm\frac{
n_\alpha-n_\beta}{2(n_\alpha n_\beta)^{1/2}}\left\{1-\frac{
n_{\stackrel{\beta}{\alpha}}^2+n_\alpha n_\beta}{P(P-
n_{\stackrel{\beta}{\alpha}}^2)}
\cos^2\theta\right\}
\ee
and from ~(\ref{7.6}) and ~(\ref{7.2a})
\be\label{7.8}
\Gamma_{\alpha,\beta}=\frac{1}{4\sqrt{n_\alpha n_\beta}\cos\theta}\left(1-
\frac{n_\alpha}{n_\beta}\right)\frac{d}{dz}(n_\beta\cos\theta).
\ee
\nl
When we consider decoupled modes, i.e.\ cases where nowhere mode coupling
occures, for the last equation follows $\Gamma_{\alpha\alpha}=\Gamma_{\beta
\beta}=0$,
and then from ~(\ref{5.8}) follows that the scalar part of the amplitude (but
not the polarization vector) does not change when the wave propagates through
the medium. This is to be expected, since wave absorbtion is a kinetic effect
and does not occure in a cold plasma (see for example
Antonson \& Manheimer 1978, sec.II), which also immediately can be concluded
from the fact that the entropy can not change, when temperature is absent.
In a cold plasma, mode coupling
can occure only in the form of reflection and transmission, without absorbtion.
\nl
We investigat now reflection at a cut-off. There, we have (Swanson 1989,
ch.\ 2.1.2)
\be\label{7.9}
n_\alpha=-n_\beta=P^{1/2}\rightarrow 0.
\ee
Hence ~(\ref{6.14}) can be used, since ~(\ref{6.8}) is valid. For ~(\ref{7.7})
and ~(\ref{7.8}) follows from ~(\ref{7.9})
\ba\label{7.10}
\Upsilon_{\stackrel{1}{2}} & = &\pm i\left(\frac{\cos^2\theta}{P}-1
\right)\\
\Gamma_{\alpha\beta} & = & -\Gamma_{\beta\alpha}\,=\,\frac{i}{2P^{1/2}}
\frac{d}{dz}\ln(P^{1/2}\cos\theta)\ea
and from ~(\ref{7.8}), ~(\ref{7.2a}) and ~(\ref{6.16}) follows
for ~(\ref{6.14})
\be\label{7.11}
\frac{d}{dz}\left( \begin{array}{c} a_\alpha \\ a_\beta \end{array} \right)
\,+\,\frac{\Gamma_{\alpha\beta}}{\Upsilon_{\scriptsize 1}^2+1} \left(
\begin{array}{cc}
\Upsilon_1 & 1 \\ -1 & \Upsilon_1 \end{array} \right) \left( \begin{array}{c}
a_\alpha \\ a_\beta \end{array} \right)\,=\,0.\ee
Using Snell's law, this can be written in the form of an equation for the
difference of the energy of the incoming and reflected wave near the
cut-off
\nl
\be\label{7.12}
\frac{d}{dz}\ln\left(|a_\alpha|^2-|a_\beta|^2\right)\,=\,
\frac{-P^{3/2}}{2\sin^4\theta_0}\,\frac{dP}{dz}\ \ \ \
\mbox{for}\ \ \ \sin\theta_0 \neq 0
\ee
\nl
{}From the definition of $P$ in ~(\ref{7.1a}), it is clear that the derivative
of $P$ to $z$
is always finite, hence
\be\label{7.13}
\frac{d}{dz}\ln\left(|a_\alpha|^2-|a_\beta|^2\right)\rightarrow 0 \ \ \
\mbox{for}\ \ \ P\rightarrow 0.\ee
(The case $\sin\theta_0=0$ can not be calculated by the above method, since
in this case Snell's law is everywhere identical to zero. But physically, we
can suppose
$\theta_0$ to be small and let the limit $P\rightarrow0$ go before
$\theta_0\rightarrow0$. Anyhow, this problem does not occure when calculating
the S-matrix, which gives equivalent results, compare eq.\ ~(\ref{7.16})).
\nl
As to be expected, the difference of the energy of the incoming and reflected
wave is conserved at the cut-off, because there no part of the wave is
transmitted. This is of course not the case, when we are away from the cut-off,
where $P$ is finite. There the difference
of the energy of the incoming and reflected wave is not conserved anymore,
because
at each stratification surface a part of the wave is transmitted toward  the
cut-off. It is only at the cut-off, where pure reflection takes
place.
\nl
Solutions for the amplitudes $a_\alpha$ and $a_\beta$ can be obtained near the
coupling point by S-matrix methods
(Volland 1962\,a,\,b, Budden 1985, ch.\ 18.5). We
define the cut off point by $z_0$, i.e.\ $P(z_0)=0$. $z_2$ is a point above,
and $z_1$ a point below $z_0$.
The solutions near $z_0$
can then be written as
\be\label{7.14}
{\bf a}(z_2)={\bf A}(z_2,z_1)\cdot{\bf a}(z_1)
\ee
where the matrizant is defined as (Gantmacher 1960)
\be\label{7.15}
{\bf A}(z_2,z_1)={\bf I}+\int_{z_1}^{z_2} d\tau_1\,{\bf \Gamma}(\tau_1)+
\int_{z_1}^{z_2}
d\tau_1\,{\bf \Gamma}(\tau_1)\cdot\int_{z_1}^{\tau_1} d\tau_2\,{\bf \Gamma}
(\tau_2)
+\cdots.
\ee
Since $z_0$ is a cut-off, $a_\alpha(z_2)$ and $a_\beta(z_2)$ are the downgoing
and upgoing modes above $z_0$ respectively and $a_\alpha(z_1)$ and
$a_\beta(z_1)$ the upgoing and downgoing modes below $z_0$ respectively
(compare to Budden  1985, eq.\,18.20).
\nl
For our purpose, it is sufficient to calculate the matrizant up to first order.
We obtain
\be\label{7.16}\begin{array}{lll}
{\bf A}(z_2,z_1)\,= & \left[\begin{array}{ll}
1-\frac{P^{5/2}(z_2)-P^{5/2}(z_1)}{10\sin^4\theta_0} &
\frac{-i(P^{9/2}(z_2)-P^{9/2}(z_1))}{18\sin^6\theta_0}\\
   &  \\
\frac{i(P^{9/2}(z_2)-P^{9/2}(z_1))}{18\sin^6\theta_0} &
1-\frac{P^{5/2}(z_2)-P^{5/2}(z_1)}{10\sin^4\theta_0}
\end{array}\right] & \ \ \ \ \ \mbox{for}\ \ \ \sin\theta_0\neq0\\
     & & \\  &  &  \\
{\bf A}(z_2,z_1)\,= & \left[\begin{array}{ll}
1-\frac{P^{1/2}(z_2)-P^{1/2}(z_1)}{2} &
\frac{-i(P^{3/2}(z_2)-P^{3/2}(z_1))}{6}\\
    & \\
\frac{i(P^{3/2}(z_2)-P^{3/2}(z_1))}{6} &
1-\frac{P^{1/2}(z_2)-P^{1/2}(z_1)}{2}
\end{array}\right] & \ \ \ \ \ \mbox{for}\ \ \ \sin\theta_0=0.\end{array}
\ee
\nl
The diagonal elements of the matrizant are the reflection and the off-diagonal
elements the transmission coefficients (Budden 1985, eq.\,18.21,
Volland 1962\,a, eq.\,21). We see,
that at the cut-off, where $z_1,z_2\rightarrow z_0$, i.e.\ $P\rightarrow0$,
the reflection coefficients
become 1 and the transmission coefficients 0, as is to be expected.
\nl
\begin{center} {\large {\bf Appendix}} \end{center}
\nl
Equation ~(\ref{5.6a}) can be written with the help of ~(\ref{5.6b}) as
\be\label{a1}
\mbox{Det}\,\M\,=\,\Theta\,(\lambda-\lambda_\alpha)(\lambda-\lambda_\beta)
\prod_{\gamma\neq\alpha,\beta}(\lambda-\lambda_\gamma)\,=\,
\Theta(\lambda^2+B\lambda+C)\prod_{\gamma\neq\alpha,\beta}(\lambda-
\lambda_\gamma)
\ee
where $B=B(\kt;\x)$ and $C=C(\kt;\x)$ are  analytical. The
roots $\lambda_\alpha$ and $\lambda_\beta$ are
\be\label{a2}
\lambda_{\stackrel{\scriptsize\alpha}{\scriptsize\beta}}=\frac{-B\pm\sqrt{B^2-
4C}}{2}.
\ee
At the coupling point, we have
\be\label{a3}
B^2-4C=0\ee
Hence the coupling point is a branch point (for the example of a plane
stratified
medium see Budden 1985, sec.\,16.3 or Budden 1972, Budden \& Smith 1974). We
have
\be\label{a4}
\frac{\partial}{\partial\kt}
\lambda_{\stackrel{\scriptsize\alpha}{\scriptsize\beta}}
=\,-\frac{1}{2}\,\frac{\partial B}{\partial\kt}\pm\frac{1}{4}\left[
\frac{\partial}{\partial\kt}(B^2-4C)\right]\frac{1}{\lambda_\alpha-
\lambda_\beta}
\ee
from which near the coupling point follows equation ~(\ref{6.8})
\nl
\newpage
\begin{center} {\large {\bf References}} \end{center}
\nl
Antonson, T.M. \& Manheimer, W.M. 1978 {\it Phys.\,Fluids}\,{\bf 21}, 2295-2305
\nl
Bernstein, I.B. \& Friedland, L. 1983 {\it Basic Plasma Physics} I, (ed.\
M.N. Rosenbluth \& R.Z. Sagdeev), North-Holland, p.367
\nl
Budden, K.G. 1972 {\it J.\,Atmos.\,Terr.\,Phys.}\,{\bf 34}, 1909-1921
\nl
Budden, K.G. 1985 {\it The Propagation of Radio waves}, Cambrige
\nl
Budden, K.G. \& Smith, M.S. 1974 {\it Proc.\,Roy.\,Soc.\,Lond.}\,A{\bf 341},
1-30
\nl
Cairns, R.A., Holt, H., McDonald, D.C., Taylor, M.\ \& Lashmore-Davies, C.N.
1995 {\it Phys.\,Plasmas}\,{\bf 2}, p.3702
\nl
Cairns, R.A. 1991 {\it Radiofrequency Heating of Plasma}, Adam Hilger
\nl
Cairns, R.A. \& Lashmore-Davies, C.N. 1983 {\it Phys.\,Fluids}\,{\bf 26},
1268-1274
\nl
Catto, P.J., Lashmore-Davies, C.N. \& Martin, T.J. 1993 {\it Phys.\,Fluids}\,B
\,{\bf 5}, 2909-2921
\nl
Friedland, L. 1985 {\it Phys.\,Fluids}{\bf 28}, 3260-3268
\nl
Friedland, L. 1990 {\it Phys.\,Fluids}\,B\,{\bf 2}, 1204-1209
\nl
Friedland, L., Goldner, G. \& Kaufman, A.N. 1987
{\it Phys.\,Rev.\,Lett.}\,{\bf 58}, 1392-1394
\nl
Gantmacher, F.R. 1960 {\it Matrix Theory}, vol.2, Chelsea
\nl
Kaufman, A.N. \& Friedland, L. 1987 {\it Phys.\,Lett.}\,A {\bf 123}, 387-389
\nl
Kravtsov, Yu.A. 1969 {\it Sov.\,Phys.\,JETP} {\bf 28}, 769-772
\nl
Mc\,Donald, D.C., Carirns, R.A. and Lashmore-Davies, C.N. 1994 {\it
Phys. Plasmas}\,{\bf 1}, 842-849
\nl
Sabzevari, B. 1992 {\it J.\,Plasma Phys.}\,{\bf 47}, 49-60
\nl
Sabzevari, B. 1993 {\it J.\,Plasma Phys.}\,{\bf 49}, 413-427
\nl
Stix, T.H. 1992 {\it Waves in Plasmas}, American Inst. Physics
\nl
Stix, T.H. \& Swanson, D.G. 1983 {\it Basic Plasma Physics} I (ed. M.N.
Rosenbluth \& R.Z. Sagdeev), North-Holland, p.335
\nl
Suchy, K. \& Altman, C. 1975 {\it J.\ Plasma Phys.}\,{\bf 13}, 299-316
\nl
Suchy, K. \& Sabzevari, B. 1992a {\it Proceedings of URSI National Committee
Germany.
Kleinheubacher Berichte 1991}, vol.\,35, 191-206
\nl
Suchy, K. \& Sabzevari, B. 1992b {\it Proceedings of 2nd Symposium on Plasma
Dynamics: Theory and Applications, Trieste}, ed.\ M.\ Tessarotto, 147-156
\nl
Swanson, D.G. 1989 {\it Plasma Waves}, Academic Press
\nl
Tskhakaya, D.D. 1996 private communication
\nl
Volland, H. 1962\,a {\it Arch. Elektr. Ubertragung}\,{\bf 16}, 328-334
\nl
Volland, H. 1962\,b {\it J.\,Atmos.\,Terr.\,Phys.}\,{\bf 24}, 853-856
\end{document}